\documentclass[11pt,letterpaper]{article}
\usepackage{graphicx}
\usepackage{CJKutf8}
\usepackage{stfloats}
\usepackage{amsmath}
\usepackage{acl2016}
\usepackage{times}
\usepackage{latexsym}

\aclfinalcopy

\title{Spotting Rumors via Novelty Detection}

 \author{Yumeng Qin \\ International School of Software \\ Wuhan University \\  yumeng.qin@whu.edu.cn \And
         Dominik Wurzer \\ School of Informatics \\ University of Edinburgh \\ s1157979@sms.ed.ac.uk \\
         \AND
         Victor Lavrenko \\ School of Informatics \\ University of Edinburgh \\ vlavrenk@inf.ed.ac.uk \And
         Cunchen Tang \\ International School of Software \\ Wuhan University \\ cctang@whu.edu.cn}

\date{2016}

\begin{document}

\maketitle

\begin{abstract}
Rumour detection is hard because the most accurate systems operate retrospectively, only recognising rumours once they have collected repeated signals. By then the rumours might have already spread and caused harm. We introduce a new category of features based on novelty, tailored to detect rumours early on. To compensate for the absence of repeated signals, we make use of news wire as an additional data source. Unconfirmed (novel) information with respect to the news articles is considered as an indication of rumours. Additionally we introduce pseudo feedback, which assumes that documents that are similar to previous rumours, are more likely to also be a rumour. Comparison with other real-time approaches shows that novelty based features in conjunction with pseudo feedback perform significantly better, when detecting rumours instantly after their publication.
\end{abstract}

\section{Introduction}
Social Media has evolved from friendship based networks to become a major source for the consumption of news (NIST, 2008). On social media, news is decentralised as it provides everyone the means to efficiently report and spread information. In contrast to traditional news wire, information on social media is spread without intensive investigation, fact and background checking. The combination of ease and fast pace of sharing information provides a fertile breeding ground for rumours, false- and disinformation. Social media users tend to share controversial information in-order to verify it, while asking about for the opinions of their followers (Zhao et. al, 2015). This further amplifies the pace of a rumour's spread and reach. Rumours and deliberate disinformation have already caused panic and influenced public opinion. \vspace{2mm}\\The cases in Germany and Austria\footnote{http://hoaxmap.org/ on 15.02.2016} in 2016, show how misleading and false information about crimes committed by refugees negatively influenced the opinion of citizens.\\ Detecting these rumours allows debunking them to prevent them from further spreading and causing harm. The further a rumour has spread, the more likely it is to be debunked by users or traditional media (Liu et. al, 2015). However, by then rumours might have already caused harm. This highlights the importance and necessity of recognizing rumours as early as possible - preferably instantaneously. \vspace{2mm}\\Rumour detection on social media is challenging due to the short texts, creative lexical variations and high volume of the streams. The task becomes even harder if we attempt to perform rumour detection on-the-fly, without looking into the future. We provide an effective and highly scalable approach to detect rumours instantly after they were posted with zero delay. We introduce a new features category called novelty based features. Novelty based features compensate the absence of repeated information by consulting additional data sources - news wire articles. We hypothesize that information not confirmed by official news is an indication of rumours. Additionally we introduce pseudo feedback for classification. In a nutshell, documents that are similar to previously detected rumours are considered to be more likely to also be a rumour.
The proposed features can be computed in constant time and space allowing us to process high-volume streams in real-time (Muthukrishnan, 2005). Our experiments reveal that novelty based features and pseudo feedback significantly increases detection performance for early rumour detection. \vspace{2mm}\\ The contributions of this paper include: \vspace{2mm}\\\textbf{Novelty based Features}\\ We introduced a new category of features for instant rumour detection that harnesses trusted resources. Unconfirmed (novel) information with respect to trusted resources is considered as an indication of rumours.\vspace{2mm}\\ \textbf{Pseudo Feedback for Detection/Classification} \\Pseudo feedback increases detection accuracy by harnessing repeated signals, without the need of retrospective operation.
\subsection{Related Work}
Before rumour detection, scientists already studied the related problem of information credibility evaluation (Castillo et. al. 2011; Richardson et. al, 2003). Recently, automated rumour detection on social media evolved into a popular research field which also relies on assessing the credibility of messages and their sources. The most successful methods proposed focus on classification harnessing lexical, user-centric, propagation-based (Wu et. al, 2015) and cluster-based (Cai et. al, 2014; Liu et. al, 2015; Zhao et. al, 2015) features. \\Many of these context based features originate from a study by Castillo et. al (2011), which pioneered in engineering features for credibility assessment on Twitter (Liu et. al, 2015). They observed a significant correlation between the trustworthiness of a tweet with context-based characteristics including hashtags, punctuation characters and sentiment polarity. When assessing the credibility of a tweet, they also assessed the source of its information by constructing features based on provided URLs as well as user based features like the activeness of the user and social graph based features like the frequency of re-tweets. A comprehensive study by Castillo  et. al (2011) of information credibility assessment widely influenced recent research on rumour detection, whose main focuses lies upon improving detection quality.\\ While studying the trustworthiness of tweets during crises, Mendoza et. al (2010) found that the topology of a distrustful tweet's propagation pattern differs from those of news and normal tweets. These findings along with the fact that rumours tend to more likely be questioned by responses than news paved the way for future research examining propagation graphs and clustering methods (Cai et. al, 2014; Zhao et. al, 2015). The majority of current research focuses on improving the accuracy of classifiers through new features based on clustering (Cai et. al, 2014; Zhao et. al, 2015), sentiment analysis (Qazvinian et. al, 2011; Wu et. al, 2015) as well as propagation graphs (Kwon, et. al, 2013; Wang et. al, 2015).\\
Recent research mainly focuses on further improving the quality of rumour detection while neglecting the increasing delay between the publication and detection of a rumour. The motivation for rumour detection lies in debunking them to prevent them from spreading and causing harm. Unfortunately, state-of-the-art systems operate in a retrospective manner, meaning they detect rumours long after they have spread. The most accurate systems rely on features based on propagation graphs and clustering techniques. These features can only detect rumours after the rumours have spread and already caused harm. \\Therefore, researchers like Liu et. al (2015), Wu et. al (2015), Zhao et. al (2015) and Zhou et. al (2015) focus on 'early rumour-detection' while allowing a delay up to 24 hours. Their focus on latency aware rumour detection makes their approaches conceptually related to ours.  Zhao et. al (1015) found clustering tweets containing enquiry patterns as an indication of rumours. Also clustering tweets by keywords and subsequently judging rumours using an ensemble model that combine user, propagation and content-based features proved to be effective (Zhou et. al, 2015). Although the computation of their features is efficient, the need for repeated mentions in the form of response by other users results in increased latency between publication and detection. The approach with the lowest latency banks on the 'wisdom of the crowd' (Liu et. al, 2015). In addition to traditional context and user based features they also rely on clustering micro-blogs by their topicality to identify conflicting claims, which indicate increased likelihood of rumours. Although they claim to operate in real-time, they require a cluster of at least 5 messages to detect a rumour. \\\\In contrast, we introduce new features to detect rumours as early as possible - preferably instantly, allowing them to be debunked before they spread and cause harm. 
\section{Rumour Detection}
\vspace{-2mm}Rumour detection is a challenging task, as it requires determining the truth of information (Zhao et. al, 2015).
The Cambridge dictionary, defines a rumour as information of doubtful or unconfirmed truth. We rely on classification using an SVM, which is the state-of-the-art approach for novelty detection. Numerous features have been proposed for rumour detection on social media, many of which originate from an original study on information credibility by Castillo et. al (2011). Unfortunately, the currently most successful features rely on information based on graph propagation and clustering, which can only be computed retrospectively. This renders them close to useless when detecting rumours early on. We introduce two new classes of features, one based on novelty, the other on pseudo feedback. Both feature categories improve detection accuracy early on, when information is limited.
\subsection{Problem Statement}
We frame the Real-time Rumour Detection task as a classification problem that assesses a document's likelihood of becoming a future rumour at the time of its publication. Consequently, prediction takes place in real-time with a single pass over the data.\\\\
More formally, we denote by $d_t$ the document that arrives from stream $S:\{d_0, d_1, . . . d_n\}$ at time $t$. Upon arrival of document $d_t$ we compute its corresponding feature vector $f_{d,t}$. Given $f_{d,t}$ and the previously obtained weigh vector $w$ we compute the rumour score $RS_{d,t} = w^T \times f_{d,t}$. The rumour prediction is based on a fixed thresholding strategy with respect to $\theta$. We predict that message $d_t$ is likely to become a rumour if its rumour score exceeds the detection threshold $RS_{d,t} > \theta$. The optimal parameter setting for weight vector $w$ and detection threshold $\theta$ are learned on a test to maximise prediction accuracy.
\subsection{Novelty-based Features}
To increase instantaneous detection performance, we compensate for the absence of future information by consulting additional data sources. In particular, we make use of news wire articles, which are considered to be of high credibility. This is reasonable as according to Petrovic et. al (2013), in the majority of cases, news wires lead social media for reporting news. When a message arrives from a social media stream, we build features based on its novelty with respect to the confirmed information in the trusted sources. In a nutshell, the presence of information unconfirmed by the official media is construed as an indication of being a rumour. Note that this closely resembles the definition of what a rumour is.



\subsection{Novelty Feature Construction}
\vspace{-1mm}High volume streams demand highly efficient feature computation. This applies in particular to novelty based features since they can be computationally expensive. We explore two approaches to novelty computation: one based on vector proximity, the other on kterm hashing.\\ 
Computing novelty based on traditional vector proximity alone does not yield adequate performance due to the length discrepancy between news wire articles and social media messages. To make vector proximity applicable, we slide a term-level based window, whose length resembles the average social media message length, through each of the news articles. This results in sub-documents whose length resembles those of social media messages. Novelty is computed using term weighted tf-idf dot products between the social media message and all news sub-documents. The inverse of the minimum similarity to the nearest neighbour equates to the degree of novelty.\\
The second approach to compute novelty relies on kterm hashing (Wurzer et. al, 2015), a recent advance in novelty detection that improved the efficiency by an order of magnitude without sacrificing effectiveness. Kterm hashing computes novelty non-comparatively. Instead of measuring similarity between documents, a single representation of previously seen information is constructed. For each document, all possible kterms are formed and hashed onto a Bloom Filter. Novelty is computed by the fraction of unseen kterms. Kterm hashing has the interesting characteristic of forming a collective 'memory', able to span all trusted resources. We exhaustively form kterm for all news articles and store their corresponding hash positions in a Bloom Filter. This filter then captures the combined information of all trusted resources. A single representation allows computing novelty with a single step, instead of comparing each social media message individually with all trusted resources.\\\\
When kterm hashing was introduced by Wurzer et. al (2015) for novelty detection on English tweets, they weighted all kterm uniformly. We found that treating all kterms as equally important, does not unlock the full potential of kterm hashing. Therefore, we additionally extract the top 10 keywords ranked by $tf.idf$ and build a separate set of kterms solely based on them. This allows us to compute a dedicated weight for kterms based on these top 10 keywords. The distinction in weights between kterms based on all versus keyword yields superior rumour detection quality, as described in section \ref{section_featureAnalysis}. This leaves us with a total of 6 novelty based features for kterm hashing - kterms of length 1 to 3 for all words and keywords.\\\\ Apart from novelty based features, we also apply a range of 51 context based features. The full list of features can be found in table \ref{tab:allFeatures}. The focus lies on features that can be computed instantly based only on the text of a message to keep the latency of our approach to a minimum. Most of these 51 features overlap with previous studies (Castillo et. al, 2011; Liu et. al, 2015; Qazvinian et. al, 2011; Yang et. al, 2012; Zhao et. al, 2015). This includes features based on the presence or number of URLs, hash-tags and user-names, POS tags, punctuation characters as well as 8 different categories of sentiment and emotions.\\\\
On the arrival of a new message from a stream, all its features are computed and linearly combined using weights obtained from an SVM classifier, yielding the rumour score. We then judge rumours based on an optimal threshold strategy for the rumour score.
\subsection{Pseudo Feedback}
In addition to novelty based features we introduce another category of features - dubbed Pseudo-Feedback (PF) feature - to boost detection performance. The feature is conceptually related to pseudo relevance feedback found in retrieval and ranking tasks in IR. The concept builds upon the idea that documents, which reveal similar characteristics as previously detected rumours are also likely to be a rumour. During detection, feedback about which of the previous documents describes a rumour is not available. Therefore, we rely on 'pseudo' feedback and consider all documents whose rumour score exceeds a threshold as true rumours.\\ The PF feature describes the maximum similarity between a new document and those documents previously considered as rumour. Similarities are measured by vector proximity in term space. Conceptually, PF passes on evidence to repeated signals by increasing the rumour score of future documents if they are similar to a recently detected rumour. Note that this allows harnessing information from repeated signals without the need of operating retrospectively. \\\\
\textbf{Training Pseudo Feedback Features}\\
The trainings routine differs from the standard procedure, because the computation of the PF feature requires two training rounds as we require a model of all other features to identify 'pseudo' rumours. In a first training round a SVM is used to compute weights for all features in the trainings set, except the PF features. This provides a model for all but the PF features. Then the trainings set is processed to computing rumour scores based on the model obtained from our initial trainings round. This time, we additionally compute the PF feature value by measuring the minimum distance in term space between the current document vector and those previous documents, whose rumour score exceeds a previously defined threshold. Since we operate on a stream, the number of documents previously considered as rumours grows without bound. To keep operation constant in time and space, we only compare against the k most recent documents considered to be rumours. Once we obtained the value for the PF feature, we compute its weight using the SVM. The combination of the weight for the PF feature with the weights for all other features, obtained in the initial trainings round, resembles the final model.
\begin{table*}[t]
\centering
\begin{tabular}{|c|l|}
\hline
\textbf{Topic Name} & \textbf{Rumour synopsis} \\ \hline
Terror in Paris & fake news report of Chinese hostages and their cruel torture \\
& by terrorist disguised as Syria refugees \\ \hline
RUS plane crash & various different rumours blaming Turkey, US and Russia itself \\ \hline
Samsung vs Apple  & rumors that Samsung paid Apple the fine with\\
Patent law suit &  trucks of coins to show their dissatisfaction with the judgement \\ \hline
Flight MH370 & rumours about the cause of the crash including abduction   \\ 
& by terrorists, shot down by US military and CIA \\ \hline
\end{tabular}
\caption{Excerpt of topics with synopsis of corresponding rumours}
\label{tab_topicList}
\end{table*}
\section{Experiments}
\vspace{-1.5mm}The previous sections introduced two new categories of features for rumour detection. Now we test their performance and impact on detection effectiveness and efficiency. In a streaming setting, documents arrive on a continual basis one at a time. We require our features to compute a rumour-score instantaneously for each document in a single-pass over the data. Messages with high rumour scores are considered likely being rumours. The classification decision is based on an optimal thresholding strategy based on the trainings set. 
\subsection{Evaluation metrics}
We report accuracy to evaluate effectiveness, as is usual in the literature (Zhou et. al, 2015). Additionally we use the standard TDT evaluation procedure (Allan et. al, 2000; NIST, 2008) with the official TDT3 evaluation scripts (NIST, 2008) using standard settings. This procedure evaluates detection tasks using Detection Error Trade-off (DET) curves, which show the trade-off between miss and false alarm probability. By visualizing the full range of thresholds, DET plots provide a more comprehensive illustration of effectiveness than single value metrics (Allan et. al, 2000). We also evaluate the efficiency of computing the proposed features, measured by the throughput per second, when applied to a high number of messages.
\subsection{Data set}
Rumour detection on social media is a novel research field without official data sets. Since licences agreements forbid redistribution of data, no data sets from previous publications are available. We therefore followed previous researchers like Liu et. al (2015) and Yang et. al (2012) and created our own dataset. \vspace{2mm}\\
\textbf{trusted resources}: We randomly collected 200 news articles about broad topics commonly reported by news wires over our target time period. These range from news about celebrities and disasters to financial and political affairs as seen in table \ref{tab_topicList}. Since we operate on Chinese social media, we gathered news articles from Xinhua News Agency\footnote{http://www.xinhuanet.com/}, the leading news-wire in China. To ensure a fair evaluation, we collected the news articles before judging rumours, not knowing which rumours we would find later on. We also only consider news articles published before the timestamps of the social media messages.\vspace{2mm}\\
For our social media stream, we chose Sina Weibo, a Chinese social media service with more than 200 million active users\footnote{http://www.bbc.co.uk/news/technology-35361157 as of 10.02.2013}. Micro-blogs from Sina Weibo are denoted as '\textit{weibos}'.\vspace{2mm}\\
\textbf{rumours}: Sina Weibo offers an official rumour debunking service, operated by trained human professionals. Following Yang et. al (2012) and Zhou et. al (2015), we use this service to obtain a high quality set of 202 confirmed rumours.\vspace{2mm}\\
\textbf{non-rumours}: We additionally gathered 202 non-rumours using the public Sina Weibo API\footnote{http://open.weibo.com/wiki/API}. Three human annotators judged these weibos based on unanimous decision making to ensure that they  don't contain rumours.\\
Since we operate in a streaming environment, all weibos are sorted based on their publication time-stamp. Table \ref{tab_rumourList} shows a list of example for rumours found in our data set. \vspace{3mm}\\
We ordered the rumours and non-rumours chronologically and divided them in half, forming a training and test set. We ensured that each of the sets consists of 50\% rumours and non-rumours. This is important when effectiveness is measured by accuracy. All training and optimization use the trainings set. Performance is then reported based on a single run on the test set.
\begin{table}[b]
\centering
\begin{tabular}{|c|c|c|c|}
\hline
Algorithm & \textbf{Our App.} & \textbf{Liu} & \textbf{Yang}   \\ \hline
Accuracy & \textbf{75\%} & 62.26\% & 60.21\% \\ \hline
Difference & - & -17\% & -20\% \\ \hline
\end{tabular}
\caption{Effectiveness in comparison with two state-of-the-art baselines for instant rumour detection using optimal thresholds}
\label{tab_eff}
\end{table}

\begin{table*}[t]
\centering
\begin{tabular}{|l|}
\hline
\begin{CJK}{UTF8}{gbsn}马航MH370最新消息，遭恐怖分子劫持过程曝光，点击查看 \end{CJK} \\ 
MH370 latest news, the process of being hijacked by terrorist has been revealed, click to view details \\ \hline
\begin{CJK}{UTF8}{gbsn}重大新闻，苹果公司获得三星10亿美元赔款，整整30卡车硬币开到了苹果公司的总部！！！ \end{CJK}  \\ 
Breaking news, Apple just got the 1 billion dollars for reparation from Samsung, \\
Samsung drove 30 trucks of coins to the headquarters of Apple. \\ \hline
\begin{CJK}{UTF8}{gbsn}法对叙IS第一轮战术核打击定于16日上午10时进行。 \end{CJK} \\ 
France will start the first round nuclear attack against IS at 10:00 a.m. on the 16th \\ \hline
\end{tabular}
\caption{Examples rumours with translation}
\label{tab_rumourList}
\end{table*}
\begin{figure}[] 
\includegraphics[width=0.48\textwidth]{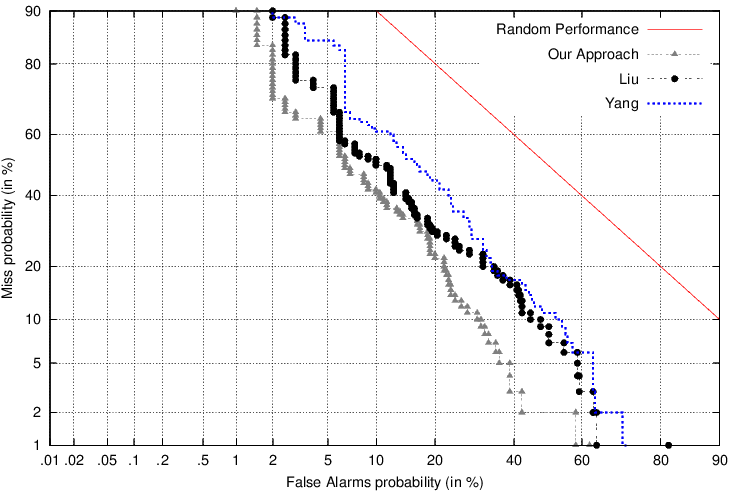}
\caption{DET plot, revealing superior effectiveness of our approach for instant rumour detection for the full range of thresholds}
\label{fig:DETplot}
\end{figure}
\begin{table}[]
\centering
\label{tab:delay}
\begin{tabular}{|c|c|c|c|}
\hline
\textbf{Algorithm} & \textbf{instant} & \textbf{12 hours} & \textbf{24 hours} \\ \hline
our App. & \textbf{75\% *} & \textbf{75\%} & 75\% \\ \hline
Liu & 62.27\% & 74.29\% & \textbf{78.4\%} \\ \hline
Yang & 60.21\% & 65.35\% & 75.82\% \\ \hline
\end{tabular}
\caption{Detection accuracy at different levels of delay; Asterisk indicates significance $(p < 0.05)$}
\end{table}
\subsection{Rumour detection effectiveness}
To evaluate our new features for rumour detection, we compare them with two state-of-the-art early rumour detection baselines Liu et. al (2015) and Yang et. al (2012), which we re-implemented. We chose the algorithm by Yang et. al (2012), dubbed Yang, because they proposed a feature set for early detection tailored to Sina Weibo and were used as a state-of-the-art baseline before by Liu et. al (2015). The algorithm by Liu et. al (2015), dubbed Liu, is said to operate in real-time and outperformed Yang, when only considering features available on Twitter. Both apply various message-, user-, topic- and propagation-based features and rely on an SVM classifier which they also found to perform best. The approaches advertise themselves as suitable for early or real-time detection and performed rumour detection with the smallest latency across all published methods. Yang performs early rumour detection and operates with a delay of 24 hours. Liu is claimed to perform in real-time while, requiring a cluster of 5 repeated messages to judge them for rumours. Note that although these algorithm are state-of-the-art for detecting rumours as quickly as possible, they still require a certain delay to reach their full potential.\\Table \ref{tab_eff} compares the performance of our features with the two classifiers on the 101 rumours and 101 non-rumours of the test set, when detecting rumour instantly after their publication. The table reveals comparable accuracy for Yang and Liu at around 60\%. Our observed performance of Yang matches those by Liu et. al (2015). Surprisingly, the algorithm Liu does not perform significantly better than Yang when applied to instantaneous rumour detection although they claimed to operate in real-time. Liu et. al (2015) report performance based on the first 5 messages which clearly outperforms Yang for early rumour detection. However, we find that when reducing the set from 5 to 1, their superiority is only marginal. In contrast, the combination of novelty and pseudo relevance based features performs significantly better (sign test with $p < 0.05$) than the baselines for instantaneous rumour detections. Novelty based features benefit from news articles as an external data source, which explains their superior performance. In particular for instantaneous rumour detection, where information can only be obtained from a single message, the use of external data proves to perform superior. Note that accuracy is a single value metric describing performance at an optimal threshold. Figure \ref{fig:DETplot} compares the effectiveness of the three algorithms for the full range of rumour scores for instantaneous detection. Different applications require a different balance between miss and false alarm. But the DET curve shows that Liu’s method would be preferable over Yang for any application. Similarly, the plot reveals that our approach dominates both baselines throughout all threshold settings and for the high-recall region in particular.\\
When increasing the detection delay to 12 and 24 hours, all three algorithms reach comparable performance with no statistically significant difference, as seen in table 4. For our approach, none of the features are computed retrospectively, which explains why the performance does not change when increasing the detection delay. The additional time allows Liu and Yang to collect repeated signals, which improves their detection accuracy. After 24 hours Liu performs the highest due to its retrospectively computed features. Note that after 24 hours rumours might have already spread far through social networks and potentially caused harm.
\begin{table*}[bp]
\centering
\begin{tabular}{|c|c|c|c|c|c|c|c|c|}
\hline
features: & \textbf{all} & \textbf{sentence char} & \textbf{POS} &  \textbf{emotion} & \textbf{extreme words} & \textbf{sentiment} & \textbf{PF} & \textbf{novelty}  \\ \hline
accuracy: & 75\% & 69\%  & 71\% & 73\% &  72\% & 73\% & 71\% & \textbf{60\%} \\ \hline
\end{tabular}
\caption{\textbf{Features ablation}: impact on performance when removing feature groups. \textbf{Note}: lower accuracy means higher impact; POS: part of speech; PF: pseudo-feedback;}
\label{tab_feature}
\end{table*}
\subsection{Feature analysis}
\label{section_featureAnalysis}
We group our 57 features into 7 categories shown in Table 6 and analyse their contribution using feature ablation, as seen in Table \ref{tab_feature}. Feature ablation illustrates the importance of a feature by measuring performance, when removing it from the set of features. Novelty related features based on kterm hashing were found to be dominant for instantaneous rumour detection $(p < 0.05)$. 'Sentence char' features, which include punctuation, hashtags, user-symbols and URLs, contributed the most of the traditional features, followed by Part of Speech ('POS') and 'extreme word' features. Our experiments found 'sentiment' and 'emotion' based features to contribute the least. Since excluding them both results in a considerable drop of performance we conclude that they capture comparable information and therefore compensated for each other.\vspace{2mm}\\
\textbf{Novelty based Features}\\
Novelty based features revealed the highest impact on detection performance. In particular kterms formed from the top keywords contribute the most. This is interesting, as when kterm hashing was introduced (Wurzer et. al, 2015), all kterms were considered as equally important. We found that prioritising certain kterms yields increased performance.\\
Interestingly, novelty based features computed by the vector similarity between weibos and news sub-documents perform slightly worse (-2\% absolute). When striping all but the top tf-idf weighted terms from the news sub-documents, the hit in performance can be reduced to -1 \% absolute. Kterm constructs a combined memory of all information presented to it. Pulling all information into a single representation bridges the gab between documents and allows finding information matches within documents. We hypothesize that this causes increased detection performance.\vspace{2mm}\\\textbf{Pseudo Feedbaack}\\Features ablation revealed that pseudo feedback (PF) increased detection performance by 5.3\% (relative). PF builds upon the output of the other features. High performance of the other features results in higher positive impact of PF. We want to further explore the behaviour of PF when other features perform badly in future studies.
 \begin{table*}[hbp]
\centering
\begin{tabular}{|c|l|}
\hline
\textbf{Category} & \textbf{Description} \\ \hline
Punctuation &  categorical feat. based on the number of !?.,\\ \hline
POS & categorical feat. based on the number of verbs, nouns, adjectives, quantity and time words \\ \hline
Sentiment & categorical feat. based on the number of strong/weak negative/positive words \\ \hline
Emotion & degree of positive/negative/sad/anxious/surprised emotion  \\ \hline
Social Media & categorical feat. based on the number of hash-tags and user-names \\ \hline
Length & categorical feat. based on the number of unique words \\ \hline
URLs & categorical feat. based on the number of URLs, pictures \\ \hline
Novelty & novelty score based kterm length 1-3 for all and key-words \\ \hline
Pseudo Feedback & distance to the closest previous rumour \\ \hline
\end{tabular}
\caption{Description of features}
\label{tab:allFeatures}
\end{table*}
\subsection{Detecting unpopular rumours}
Previous approaches to rumour detection rely on repeated signals to form propagation graphs or clustering methods. Beside causing a detection delay these methods are also blind to less popular rumours that don't go viral. In contrast, novelty based feature require only a single message enabling them to detect even the smallest rumours. Examples for such small rumours are shown in table \ref{tab_rumourList}. \vspace{-1mm}
\begin{figure}[h] 
\includegraphics[width=0.48\textwidth]{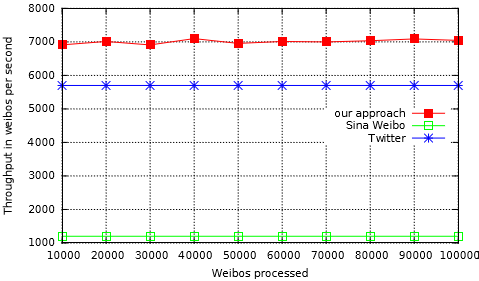}
\caption{Throughput of our approach per second in comparison to the average Twitter (Firehose) and Sina Weibo stream}
\label{fig:throughput}
\end{figure}
\vspace{-3mm}
\subsection{Efficiency and Scalability}
To demonstrate the high efficiency of computing novelty and pseudo feedback features, we implement a rumour detection system and measure its throughput when applied to 100k weibos. We implement our system in C and run it using a single core on a 2.2GHz Intel Core i7-4702HQ. We measure the throughput on an idle machine and average the observed performance over 5 runs. Figure \ref{fig:throughput} presents performance when processing more and more weibos. The average throughput of our system is around 7,000 weibos per second, which clearly exceeds the average volume of the full Twitter\footnote{about.twitter.com/company (last updated: 31.3.2015)} (5,700 tweets/sec.) and Sina Weibo\footnote{http://www.techweb.com.cn/internet/2012-01-06/1139327.shtml} (1,200 weibos/sec.) stream. Since the number of news articles is relatively small, we find no difference in terms of efficiency between computing novelty features based on kterm hashing and vector similarity. Figure \ref{fig:throughput} also illustrates that our proposed features can be computed in constant time with respect to the number of messages processed. This is crucial to keep operation in a true streaming environment feasible. Approaches, whose runtime depend on the number of documents processed become progressively slower, which is inapplicable when operating on data streams. Our experiments show that the proposed features perform effectively and their efficiency allows them to detect rumours instantly after their publication. 
\section{Conclusion}
We introduced two new categories of features which significantly improve instantaneous rumour detection performance. Novelty based features consider the increased presence of unconfirmed information within a message with respect to trusted sources as an indication of being a rumour. Pseudo feedback features consider messages that are similar to previously detected rumours as more likely to also be a rumour. Pseudo feedback and its variant, recursive pseudo feedback, allow harnessing repeated signals without the need of operating retrospectively. Our evaluation showed that novelty and pseudo feedback based features perform significantly more effective than other real-time and early detection baselines, when detecting rumours instantly after their publication. This advantage vanishes when allowing an increased detection delay. We also showed that the proposed features can be computed efficiently enough to operate on the average Twitter and Sina Weibo stream while keeping time and space requirements constant.\\\\\\\\\\

\bibliographystyle{plain}

\begin{thebibliography}{plainnat.bst}
\bibitem{fllf99}
James Allan, Victor Lavrenko and Hubert Jin. First story detection in TDT is hard. In Proceedings of the ninth international conference on Information and knowledge management. ACM, 2000
\bibitem{fkkkf9}
James Allan. Topic Detection and Tracking: Event-Based Information Organization. Kluwer Academic Publishers, Norwell, 2002
\bibitem{aaf2}
G. Cai, H. Wu, R. lv, rumours Detection in Chinese via Crowd Responses, ASONAM 2014, Beijing, China, 2014
\bibitem{ff2}
Castillo C, Mendoza M, Poblete B. Information Credibility On Twitter[C], The 20th International Conference on World Wide Web, Hyderabad, India, 2011
\bibitem{vvmml}
S. Kwon, M. Cha, K. Jung, W. Chen and Y. Wang, "Prominent features of rumor propagation in online social media", Data Mining (ICDM),  IEEE, 2013
\bibitem{f3f2}
X. Liu, A. Nourbakhsh, Q. Li, R. Fang, and S. Shah, Real-time rumor debunking on twitter, in Proceedings of the 24th ACM International Conference on Information and Knowledge Management. ACM, 2015
\bibitem{nmr89}
M. Mendoza, Poblete B, Castillo C, Twitter Under Crisis: Can we Trust What we RT?, The 1st Workshop on Social Media Analytics, SOMA, 2010
\bibitem{f22}
S. Muthukrishnan. Data streams: Algorithms and applications. Now Publishers Inc, 2005
\bibitem{assvdavsdf3}
S. Petrovic, M. Osborne, R. McCreadie, C. Macdonald, I. Ounis, and L. Shrimpton. Can Twitter replace Newswire for breaking news? In Proc. of ICWSM, 2013
\bibitem{ff2s}
V. Qazvinian, E. Rosengren, D. R. Radev, Q. Mei, rumour has it: Identifying Misinformation in Microblogs, EMNLP, July 27-31, 2011, Edinburgh, UK, 2011
\bibitem{fs}
Richardson M, Agrawal R, Domingos P, Trust Management for the Semantic Web; Lecture Notes in Computer Science, 2003: 351-368.
\bibitem{fsdddd222}
Fensel D, Sycara K, Mylopoulos J, The Semantic Web-ISWC 2003, Heidelberg: Springer BerlinHeidelberg, 2003
\bibitem{aaddc2}
S. Sun, H. Liu, J. He, X, Du, Detecting Event rumours on Sina Weibo Automatically, APWeb 2013
\bibitem{fjfjsjsjs9}
TDT by NIST - 1998-2004. http://www.itl.nist.gov/iad/mig/tests/tdt/resources.html (Last Update: 2008)
\bibitem{fls0}
K. Wu, S. Yang, K. Zhu, False rumours Detection on Sina Weibo by Propagation Structures, In the Proceedings of ICDE, 2015
\bibitem{ssssjjjjj}
Shihan Wang, Takao Terano, Detecting rumour patterns in streaming social media, Guimi, IEEE, 2015
\bibitem{domdom1}
Dominik Wurzer, Victor Lavrenko, Miles Osborne. Tracking unbounded Topic Streams. In the Proceedings of the 53rd Annual Meeting of the Association for Computational Linguistics, ACL, 2015
\bibitem{domdom}
Dominik Wurzer, Victor Lavrenko, Miles Osborne. Twitter-scale New Event Detection via K-term Hashing. In the Proceedings of the  Conference on Empirical Methods in Natural Language Processing, EMNLP, 2015
\bibitem{assvdavsdf}
Fan Yang, Xiaohui Yu, Yang Liu, Min Yang (2012) Automatic Detection of Rumor on Sina Weibo. MDS’12. August 12, 2012
\bibitem{f22da}
Z. Zhao, P. Resnick, and Q. Mei Enquiring Minds: Early Detection of Rumors in Social Media from Enquiry Posts. In the Proceedings of WWW, 2015
\bibitem{f22dafff}
X. Zhou, J. Cao, Z. Jin, X. Fei, Y. Su, J. Zhang, D. Chu, and X Cao. Realtime news certification system on sina weibo. WWW 2015
\end{thebibliography}

\end{document}